\documentclass[12pt]{article}
\pdfoutput=1

\newlength{\abstractwidth}
\usepackage{amsmath}
\usepackage{amsfonts}
\usepackage{amssymb}
\usepackage{latexsym}
\usepackage{cite}

\usepackage{epsf}
\usepackage{color}
\usepackage{graphicx}
\usepackage{dsfont}

\renewcommand{\thefootnote}{\fnsymbol{footnote}}
\renewcommand{\thanks}[1]{\footnote{#1}}
\newcommand{\starttext}{
\setcounter{footnote}{0}
\renewcommand{\thefootnote}{\arabic{footnote}}}

\newcommand{\bea}{\begin{eqnarray}}
\newcommand{\eea}{\end{eqnarray}}
\newcommand{\be}{\begin{eqnarray}}
\newcommand{\ee}{\end{eqnarray}}
\newcommand{\bma}{\begin{matrix}}
\newcommand{\ema}{\end{matrix}}
\newcommand{\<}{\langle}
\renewcommand{\>}{\rangle}

\def\cO{{\cal O}}

\def\cW{{\cal W}}

\def\bw{{\bf w}}

\def\bz{{\bf z}}

\def\ZZ{{\mathbb Z}}
\def\RR{{\mathbb R}}

\def\CC{{\mathbb C}}

\def\tr{{\rm tr}}

\def\half{{1\over 2}}

\def\p{\partial}

\def\eps{\epsilon}

\def\ep{\varepsilon}
\def\om{\omega}

\def\no{\nonumber}
\def\sm{\smallskip}

\def\Lc{{\cal L}}

\begin{document}
\starttext
\setcounter{footnote}{0}

\begin{flushright}
2019 December 6 \\
\end{flushright}

\vskip 0.3in

\begin{center}

{\Large \bf Gravitational Wilson lines in AdS$_{\bf 3}$
\footnote{Research supported in part by the National Science Foundation under grant PHY-19-14412.}}

\vskip 0.3in

{\large  Eric D'Hoker and Per Kraus}

\vskip 0.1in

{ \sl Mani L. Bhaumik Institute for Theoretical Physics}\\
{\sl Department of Physics and Astronomy }\\
{\sl University of California, Los Angeles, CA 90095, USA}

\vskip 0.05in

{\tt \small dhoker@physics.ucla.edu, pkraus@ucla.edu}

\begin{abstract}
The construction of gravitational Wilson lines in the Chern-Simons formulation of $AdS_3$ gravity  in terms of composite operators in the dual boundary conformal field theory is reviewed. New evidence is presented that the Wilson line, dimensionally regularized and suitably renormalized,  behaves as a bi-local operator of two conformal primaries whose dimension is predicted by $SL(2,\RR)$ current algebra.
\end{abstract}

\vskip 0.3in

{\sl Dedicated to Professor Roman Jackiw on the occasion of his 80-th birthday}

\end{center}


\baselineskip=15pt
\setcounter{equation}{0}
\setcounter{footnote}{0}

\newpage

\section{Introduction}
\setcounter{equation}{0}
\label{sec:1}

The deep impact of Roman Jackiw's contributions to quantum field theory extends from particle physics and condensed matter physics to mathematics. The fundamental role played by the Adler-Bell-Jackiw anomaly \cite{Bell:1969ts,Adler:1969gk} and its generalizations  in the renormalizability of Yang-Mills theory \cite{Gross:1972pv,Bouchiat:1972iq},  the non-perturbative dynamics of gauge theories \cite{tHooft}, and the consistency of string theory \cite{AlvarezGaume:1983ig,Green:1984sg} is well-known.  Conformal symmetry \cite{Callan:1970ze}, Chern-Simons field theory \cite{Deser:1981wh,Jackiw:1990mb}, and three-dimensional gravity \cite{Deser:1983nh} are but a few other subjects to which Roman has contributed brilliantly, and which provide the arena for the present work on quantum observables in $AdS_3$, along with Liouville theory  \cite{DHoker:1982wmk} in collaboration with the first author, and two-dimensional gravity \cite{Jackiw:1984je,Teitelboim:1983ux}.

\sm

While three-dimensional gravity does not support propagating gravitational waves or gravitons, it does have black hole solutions of finite mass, Bekenstein-Hawking entropy and temperature \cite{Banados:1992wn}. Accounting for the entropy of these black holes in terms of a precise counting of quantum micro-states should provide a simplified but informative warm-up for the study of the quantum behavior of  physical four-dimensional black holes.

\sm

Gravity in a space-time which is  asymptotically $AdS_3$ may be reformulated in terms of Chern-Simons theory for gauge group $SL(2,\RR) \times SL(2,\RR)$ \cite{Achucarro:1987vz,Witten:1988hc}. Classical solutions, such as thermal AdS and BTZ black holes, then correspond to flat connections, and $AdS_3$ gravity is now power-counting renormalizible \cite{Witten:1988hc}. Furthermore, $AdS_3$ gravity is holographically dual to two-dimensional conformal field theory, where powerful methods  are available for counting states and evaluating correlation functions. In fact, an early clue to the existence of AdS/CFT duality was the discovery of the Virasoro asymptotic symmetry of $AdS_3$  with a central charge proportional to the radius of $AdS_3$ \cite{Brown:1986nw}.  Whereas the large central charge behavior  is semi-classical, the regime of finite central charge corresponds to fully quantized gravity in $AdS_3$.  There is increasing evidence that $AdS_3$  may well provide the first example where the AdS/CFT duality can be proven \cite{Eberhardt:2019ywk}.

\sm

Fundamental observables in Chern-Simons theory are the Wilson line and the Wilson loop, namely  path-ordered integrals of the gauge connection  in diverse representations of the gauge group. Given the flatness of the connection on physical $AdS_3$ solutions, the (closed) Wilson loop measures the holonomy of the connection; for instance its value when wrapping the horizon of a BTZ black hole yields the Bekenstein-Hawking entropy \cite{deBoer:2013gz,Ammon:2013hba}.    Another interesting observable is the open Wilson line anchored by two points on the conformal boundary of $AdS_3$.  It is invariant under gauge transformations and diffeomorphisms that vanish at the boundary.  In gravitational language, its expectation value is related to the propagation of a massive point particle, including the effects of gravitational self-interaction which renormalize its mass \cite{Besken:2017fsj}.  The possibility of understanding such quantum gravity effects in a controlled setting provides one motivation for studying the quantum properties of the open Wilson line.

\sm

The Wilson line in the bulk of $AdS_3$  has a natural counterpart in the boundary, given by a Wilson line for a  composite gauge field built out of the stress tensor of the CFT. This observable arises naturally by using the flatness of the connection in the bulk to push the Wilson line from the bulk onto the boundary.  As reviewed below, classical considerations (i.e. large $c$) indeed confirm that this Wilson line between two points $z_1, z_2$ on the boundary corresponds to a bi-local observable in the CFT which transforms under conformal transformations as a product of two conformal primaries of identical dimensions located at the points $z_1$ and $z_2$.  More precisely, it yields the Virasoro vacuum OPE block, capturing all operators built out of the stress tensor that appear in the operator product expansion of the two primaries.   The composite nature of the corresponding quantum Wilson line operator requires subtle regularization and renormalization. Bi-locality and the  conformal properties of the Wilson line are obscured during the regularization and renormalization process.  References that develop the role of Wilson lines, and networks of Wilson lines, as representing conformal blocks include  \cite{Bhatta:2016hpz,Besken:2016ooo,Fitzpatrick:2016mtp,Besken:2017fsj,
Hikida:2018dxe,Hikida:2017ehf,Anand:2017dav,Kraus:2018zrn,Besken:2018zro}.
We also note \cite{Verlinde:1989ua} which foreshadows some of these developments.

\sm

In the present paper we shall begin by reviewing earlier work \cite{Besken:2016ooo,Hikida:2018dxe,Besken:2018zro} where an expansion in powers of $1/c$ was used to establish agreement between the quantum dimension of the Wilson line predicted from general considerations of $SL(2,\RR)$ current algebra and explicit calculations in perturbation theory in $1/c$ to order $1/c^3$ included. We shall then extend these results by showing that the correlator between a single stress tensor and the Wilson line is as predicted by conformal invariance to order $1/c$ and that the correlators between an arbitrary number of stress tensors and the Wilson line agree with predictions from conformal field theory to leading order in $1/c$, thereby providing further evidence that the Wilson line operator behaves as a bi-local operator of conformal primaries.


\section{$AdS_3$ Chern-Simons}
\setcounter{equation}{0}
\label{sec:2}

In this section we shall review the formulation of $AdS_3$ gravity in terms of $SL(2,\RR) \times SL(2,\RR)$ Chern-Simons theory.

\subsection{Chern-Simons}

We consider an oriented 3-dimensional manifold $M$ which is locally asymptotic to $AdS_3$ and whose conformal boundary is a Riemann surface $\Sigma$. In terms of a metric $g$ on $M$ the standard Einstein-Hilbert action with negative cosmological constant is given by,
\bea
 I[g] = -{ 1 \over 16 \pi G} \int _M d^3 x \sqrt{g} \left ( R + { 2 \over \ell^2} \right )
 \eea
 where $G$ is the three-dimensional Newton constant, $R$ is the Ricci scalar curvature, and $\ell>0$ is the radius of the $AdS_3$ vacuum solution to Einstein's equations. Equivalently, the action may be recast in terms of the frame $e^a$ and connection $\omega ^a$ one-forms with $a=1,2,3$ and,  as a special feature of three dimensions, without reference to the inverse frame.  A convenient way to package this data is in terms of two gauge fields  $A,  \tilde A$ obtained as a linear combination of the frame $e^a$ and the connection $\om^a$,
 \bea
 \label{Afield}
A = A^a t_a = \left (  \omega ^a + \lambda e^a \right ) t_a
\hskip 1in
\tilde A = \tilde A^a t_a=  \left (  \omega ^a - \lambda e^a \right ) \tilde t_a
\eea
whose dynamics is governed by a Chern-Simons action,
\bea
S[A, \tilde A] = -{ k \over 4 \pi} \int _M \tr \left ( A dA + \tfrac{2}{3} A^3 \right )
 +{ k \over 4 \pi} \int _M \tr \left ( \tilde A d \tilde A + \tfrac{2}{3} \tilde A^3 \right ) + S_{\p M} [A,\tilde A]
\eea
Here $S_{\p M}$ is the boundary  action required for the variational principle, as discussed  in \cite{Campoleoni:2010zq}, but whose explicit form will not be needed here.

\sm

$AdS_3$ with Minkowski signature has isometry group $SO(2,2)=SL(2,\RR) \times SL(2,\RR)/\ZZ_2$. The gauge fields $A $ and $\tilde A$ are independent of one another and take values in the Lie algebra of $SL(2,\RR) \times SL(2,\RR)$. \footnote{The generators $t_a$ and $\tilde t_a$ of the Lie algebra of $SL(2,\RR) \times SL(2,\RR)$ may be chosen  real and  normalized by $\tr (t_at_b) = \tr (\tilde t_a \tilde t_b) =  \half \eta _{ab}$ for $\eta = {\rm diag} (++-)$ with $[t_a, t_b] = \ep_{abc} \eta ^{cd} t_d$ and $[ \tilde t_a, \tilde t_b] = \ep_{abc} \eta ^{cd} \tilde t_d$ for $\ep_{123}=1$. For example, in the defining representation we may choose $t_1= \half \sigma _3$, $t_2 = \half \sigma _1$, $t_3 = - {i \over 2} \sigma _2$ in terms of Pauli matrices. We shall often prefer to use a Cartan basis of generators $L_0, L_{\pm 1}$ whose structure relations are $[L_m, L_n] = (m-n) L_{m+n}$ and which are related to the $t_a$ generators by $L_0=t_1$ and $L_{\pm 1} = t_3 \pm t_2$. \label{foot1}} The parameters $k$ and $\lambda$ are real and related to $\ell$ by $4Gk =  \ell$  and $\lambda = 1/\ell$, and the level $k$ is quantized in integer values.

\sm

$AdS_3$ with Euclidean signature is, however, the natural framework for the holographic correspondence with two-dimensional CFT on a compact conformal boundary Riemann surface~$\Sigma$. The isometry group  $SO(3,1)$   is no longer the product of rank one Lie groups.  To recover the Chern-Simons formulation, we complexify the  isometry group  $SO(3,1)$ to $SO(4,\CC) = SL(2,\CC) \times SL(2,\CC)/\ZZ_2$ and complexify the frame $e^a$ and the connection $\om^a$. The gauge fields $A, \tilde A$ are still given by (\ref{Afield}) but now with imaginary $\lambda$,
\bea
\lambda = { i \over \ell} \hskip 1ink = {\ell \over 4G}
\eea
The generators $t_a$ and $\tilde t_a$ of the Lie algebra of $SL(2,\CC) \times SL(2,\CC)$ may be chosen to coincide with those of $SL(2,\RR) \times SL(2,\RR)$ as given in the footnote, but the Lie algebra is now over~$\CC$. In this complexified formulation, the fields $A$ and $\tilde A$ are independent of one another, just as was the case with Minkowski $AdS_3$. Physical solutions, for which $e^a$ and $\om^a$ must be real,  are obtained by imposing the condition $\tilde A^a = - (A^a)^*$. For the holographic correspondence with the boundary CFT it will be convenient to work with the complexified formulation to obtain chiral conformal blocks, and use the reality condition only to construct a Hermitian pairing of left and right-moving conformal blocks.

\sm

The field equations in the bulk express the flatness of both connections,
\bea
F = dA + A \wedge A =0 \hskip 1in \tilde F = d \tilde A + \tilde A \wedge \tilde A =0
\eea
which are solved locally by $ A = U^{-1} d U$ and $\tilde A = \tilde U^{-1} d \tilde U$, for $U, \tilde U \in SL(2,\CC)$.

\subsection{$AdS_3$ asymptotics}

Metrics on the three-dimensional space-time $M$ which are asymptotically $AdS_3$ may be parametrized by Fefferman-Graham coordinates $(r, x^\mu)$ for $ \mu =1,2$. The coordinate $r$ is transverse to the conformal boundary $\Sigma$ (which is reached in the limit  $ r \to \infty$)  and $x^\mu$ are local  coordinates parallel to the boundary. The metric $ds^2$ then takes the form,
\bea
\label{metric}
{ 1 \over \ell^2} \, ds^2 = { dr^2 \over r^2} + r^2 \gamma _{\mu \nu} (r, x) dx^\mu dx ^\nu
\eea
The transverse metric $\gamma _{\mu \nu} $ admits an expansion in powers of $r^2$ for large $r$ given by,
\bea
\gamma _{\mu \nu} (r, x) =
\gamma ^{(0)} _{\mu \nu }(x)  + { 1 \over r^2} \gamma ^{(2)}_{\mu \nu } (x)  + \cO(r^{-4})
\eea
With a suitable gauge choice for the radial component of the gauge fields,
\bea
A_r (r,x) = {1 \over r}  L_0 \hskip 1in \tilde A_r (r,x) = -{ 1 \over r} \tilde L_0
\eea
Fefferman-Graham coordinates may also be used to express the gauge fields,
\bea
A (r,x) & = & + \, { d r \over r} L_0 + r A^{(0)}(x) + A^{(1)}(x)  + \cO(r^{-1})
\no \\
\tilde A (r,x)  & = & - \, { d r \over r} \tilde L_0 + r \tilde A^{(0)}(x)  + \tilde A^{(1)}(x) + \cO(r^{-1})
\eea
where $A^{(0)}, \tilde A^{(0)}, A^{(1)}, \tilde A^{(1)}$ have vanishing components along the differential $dr$. According to the standard AdS/CFT dictionary, $\gamma ^{(0)}, A^{(0)},$ and $\tilde A^{(0)}$ are the sources to the bulk fields, while  $\gamma ^{(2)}, A^{(1)},$ and $\tilde A^{(1)}$ are the expectation values of the dual CFT operators; see, e.g., \cite{Kraus:2006wn,Campoleoni:2010zq}.

\sm

We shall be interested in pure gravity solutions with $AdS_3$ asymptotics. Locally, we can set $\gamma^{(0)}$ equal to the  flat metric on $\Sigma$.  The metric endows $\Sigma$  with a complex structure, and we choose local complex coordinates $z, \bar z$ in terms of which the metric is given by,
\bea
\gamma _{\mu \nu} ^{(0)} dx^\mu dx ^\nu = dz \, d \bar z
\eea
The boundary conditions on $A$ and $\tilde A$ depend on the complex structure, and are as follows,
\bea
A^{(0)} (x) = A^{(0)} _z (z) dz \hskip 1in \tilde A^{(0)} (x) = \tilde A^{(0)} _{\bar z} (\bar z) d \bar z
\eea
where the flatness of the connections implies that $A^{(0)} _z (z)$ is holomorphic in $z$ and $\tilde A^{(0)} _{\bar z} (\bar z)$ is holomorphic in $\bar z$. The $\gamma _{\mu \nu}^{(2)}$ part of the metric is related to the expectation value of the stress tensor  $T$ which, in the above complex coordinates, is given as follows,
\bea
\gamma _{\mu \nu}^{(2)} dx^\mu dx^\nu = { 6 \over c} \Big ( T (z) dz^2 + \tilde T (\bar z)  d\bar z^2 \Big )
\eea
Since the boundary theory is conformal, the trace part of the stress tensor is absent \cite{Kraus:2006wn}. The asymptotic form of the metric (\ref{metric}) to order $\cO(r^{-2})$ is invariant under infinitesimal conformal transformations $\delta z= \ep(z)$ (accompanied by $2 \delta \rho = - \p_z \ep$ and $2 r^2 \delta \bar z = -  \p_z^2 \ep$) provided $T$ transforms as the stress tensor of a CFT of central charge $c$, whose value in terms of $G$ and $\ell$ is  given by the Brown-Henneaux formula \cite{Brown:1986nw},
\bea
\label{deltaT}
\delta T = \ep \p _z T + 2 (\p_z \ep) T - { c \over 12} \p_z^3 \ep
\hskip 1in
c= { 3 \ell \over 2G}
\eea

\subsection{The general solutions asymptotic to $AdS_3$}

The general solution takes the form,
\bea
\label{Asol}
A = b(r) ^{-1} \Big ( d + a (z)  \Big ) b(r) & \hskip 0.6in & a= dz\, L_1  + { 6 \over c} T(z) dz \, L_{-1}
\no \\
\tilde A = \tilde b(r)  \Big ( d + \tilde a (\bar z)  \Big ) \tilde b(r)^{-1} &&
\tilde a = d\bar z \, \tilde L_1 + { 6 \over c} \tilde T( \bar z) d\bar z \, \tilde L_{-1}
\eea
where $b(r) = r^{L_0} $ and $\tilde b (r) = r^{ \tilde L_0}$.
The corresponding metric is given by,
\bea
{ 1 \over \ell^2} ds^2 = { dr ^2 \over r^2} +  r^2|dz|^2
+ { 6 \over c} \left ( T(z) dz^2 + \tilde T(\bar z) d\bar z^2 \right )
+ { 36  \over c^2 \, r^2} T(z) \tilde T(\bar z) |dz|^2
\eea
The solution is exact for any holomorphic $T(z)$ in $z$  and $\tilde T(\bar z)$ in $\bar z$. The real solution is obtained by setting $ \tilde A^a = - (A^a)^*$ and  $\tilde T (\bar z) = T(z)^*$.


\section{Gravitational Wilson lines}
\setcounter{equation}{0}
\label{sec:3}

In this section, we shall review gravitational Wilson lines and their role as bi-local conformal primary fields in the holographic dual conformal field theory.

\subsection{The classical  Wilson line}

We begin by considering the Wilson line in the background gauge field of a general classical solution given in the preceding section. Henceforth, we shall restrict attention to the chiral sector of the theory, which is governed by the chiral stress tensor $T$ and the chiral gauge field $A$ which takes values in a representation of the Lie algebra of $SL(2,\CC)$ labelled by its spin $j$. As proposed in \cite{Besken:2016ooo,Besken:2017fsj} it suffices to consider the finite-dimensional representations of $SL(2,\CC)$ for which $j$ is a positive half-integer. The classical Wilson line $\cW [Z_2, Z_1]$ between two arbitrary points $Z_i = (r_i, z_i)$ for $i=1,2$ in the bulk, is defined as follows,
\bea
\cW_A [Z_2, Z_1] = P \exp \int _{Z_1} ^{Z_2} A
\eea
Path ordering is required even classically because $A$ takes values in the non-Abelian Lie algebra of $SL(2,\CC)$.
The Wilson line $\cW_A$ takes values in the representation of the group $SL(2,\CC)$ of spin $j$.
Under a gauge transformation $U \in SL(2,\CC)$ with $A \to U^{-1} (d +A) U$ the Wilson line transforms as follows,
\bea\label{gtW}
\cW_A [Z_2, Z_1]  \to U(Z_2)^{-1} \cW_A [Z_2, Z_1] U(Z_1)
\eea
Since all $r$-dependence of $A$ arises in (\ref{Asol}) through a gauge transformation $b$ of an $r$-independent gauge field $a$, we may extract all $r$-dependence of the Wilson line,
\bea
\cW_A[Z_2, Z_1] = b(r)^{-1} \cW_a[z_2, z_1] b(r) \hskip 1in \cW_a[z_2, z_1] = P \exp \int _{z_1} ^{z_2} a
\eea
In particular, the $r$-dependence of the matrix element of $\cW_A$ between highest and lowest weight states $|j, \pm j\>$  (which satisfy $b(r) |j,  \pm j \> = r^{\pm j} |j, \pm j\>$) has  $r$-dependence,
\bea
\<j,  -j | \cW_A [Z_2, Z_1] |j, j\> = r^{-2h} \<j,  -j | \cW_a [z_2, z_1] |j, j\>
\eea
which indicates that the classical Wilson line  has dimension $h=-j$.

\sm

Having factored out the $r$-dependence, the central object of study is the remaining matrix element for given $j$ which we shall denote by,
\bea
\label{Wilson}
W[z_2, z_1] =
\< j, -j |  P \exp \int _{z_1} ^{z_2} dz \left ( L_1 + { 6 \over c} T(z) L_{-1}  \right ) |j, j\>
\eea
The classical Wilson line $W$ transforms under local conformal transformations as a bi-local primary field of dimension $h=-j$ at both points $z_1$ and $z_2$. To see this, we start with $T=0$, in which case the path ordered integral reduces to an ordinary integral and we have,
\bea
 W[z_2, z_1] \Big |_{T=0} =
\<j,  -j | e ^{(z_2-z_1) L_1}  |j, j\>= (z_2-z_1)^{-2h}
\eea
Under an arbitrary local conformal transformation $z \to f(z)$, the function $T$ transforms to a non-zero value, given by the Schwarzian of $f$,
\bea
\label{ai}
T_f(z) =   {c\over 12} \left (  {f'''(z)\over f'(z)}-{3\over 2} \left( f''(z)\over f'(z)\right)^2 \right )
\eea
as may be established by integrating the infinitesimal transformation law of (\ref{deltaT}).
The Wilson line for this value of $T=T_f$ is found to be,
\bea
\label{aj}
 W[z_2, z_1] \Big |_{T=T_f} =  { [f'(z_2) f'(z_1)]^h \over [ f(z_2)-f(z_1)]^{2h}}
\eea
where we again have $h=-j$.   Although the path-ordered integral depends on the values of $T(z)$ along the integration path from $z_1$ to $z_2$, the value of the above matrix element of the Wilson line depends only on the end-points, so that the Wilson line indeed behaves as a bi-local conformal primary field of dimension $h$ at the points $z_1$ and $z_2$. Bi-locality may be traced back, of course, directly to the fact that the Chern-Simons field equations express the flatness of the connection $A$ in the bulk of $AdS_3$, so that the value of the Wilson line between any two points is independent of the path between the points.

\subsection{The quantum Wilson line}
\label{sec:25}

The classical Wilson line $W[z_2,z_1]$ defined in (\ref{Wilson}) may be promoted to a quantum operator in the boundary CFT by promoting $T$ to the stress tensor operator of the CFT. This construction is formal as short-distance singularities in the OPE of stress tensor operators,
\bea
\label{TT}
T(z) T(w) = { c/2 \over (z-w)^4} + { 2 T(w) \over (z-w)^2 } + { \p_w T(w) \over z-w} + \hbox{regular}
\eea
arise in the path-ordered exponential when two or more stress tensor operators collide. The presence of these short-distance singularities  will require regularization and renormalization. For finite values of $c$ the quantum Wilson line, if it can be suitably renormalized, will provide quantum observables of $AdS_3$ gravity at finite coupling, whence the importance of the construction of these operators. In the limit of large $c$, the correlators of the quantum Wilson line are expected to tends to the values predicted by the classical Wilson line.

\sm

The central question of the present investigations is whether a renormalized quantum Wilson line operator $W_R[z_2,z_1]$ can be constructed with the following properties.
\begin{enumerate}
\itemsep=0in
\item The renormalized Wilson line  $W_R[z_2,z_1]$
is equivalent to a bi-local operator $\cO(z_2) \cO(z_1)$,
in the sense that their correlators  with  an arbitrary number $p$ of stress tensors coincide,
\bea
\label{al}
\<  T(w_1) \cdots T(w_p) W_R [z_2,z_1]\> = \<  T(w_1)\cdots T(w_p) \cO(z_2) \cO(z_1) \>
 \eea
These identities mean that the Virasoro vacuum OPE block may be generated by the  operator $W_R[z_2,z_1]$, so that the renormalized Wilson line operator $W_R[z_2,z_1] $ captures all terms in the OPE of $\cO(z_2) \cO(z_1)$ which involve only the stress tensor.
\item The Wilson line $W_R$ and the operator $\cO$ have chiral conformal dimension $h$ which coincides with the predictions from $SL(2,\RR)$ current algebra \cite{Bershadsky:1989mf},
\bea
\label{hj}
h (j) = - j + { m+1 \over m} j(j+1) \hskip 1in c = 1 - { 6 \over m (m+1)}
\eea
\end{enumerate}
If a renormalized quantum operator $W_R$ with these properties can be constructed, then its vacuum expectation value must be given by,
\bea
\label{Wconf}
\< W_R[z_2,z_1] \> = (z_2-z_1)^{-2h(j)}
\eea
while its correlator with one stress tensor must be,
\bea
\label{TWconf}
\<  T(w) \, W_R[z_2,z_1]  \> = { h(j) \, (z_2-z_1)^2 \over (z_2-w)^2 (z_1-w)^2} \< W_R[z_2,z_1] \>
\eea
We note that the exact order 2 of the poles in $w$ at $z_1$ and $z_2$ of this correlator  signifies that $T(w)$ suffers no discontinuity as it is moved across the line of integration from $z_1$ to $z_2$, thereby confirming the bi-local nature of the operator $W_R$. The remainder of this paper, starting in the subsequent section, will be devoted to developing evidence for the existence of a renormalized Wilson line operator $W_R$ with these properties.

\subsection{Wilson line as solution to Virasoro Ward identity}
\label{wardsec}
\setcounter{equation}{0}

In this section we show how to arrive at the Wilson line from the perspective of Virasoro Ward identities, which are equivalent to imposing condition (1) in the previous subsection.  We show that the unrenormalized Wilson line formally solves these Ward identities.  The solution is formal because the unrenormalized Wilson line is singular as a quantum operator. However, UV divergences only set in at subleading order in $1/c$, and so the following argument will establish the validity of the Wilson line, in the sense of obeying conditions (1) and (2), at leading order in $1/c$.

We begin with a definition of Virasoro  OPE blocks \cite{Czech:2016xec}.
Let ${\cal O}_i(z)$ denote a Virasoro primary field of (chiral) scaling dimension $h_i$, where we suppress dependence on anti-holomorphic data.    Given two such primary fields we  can act on the CFT  vacuum state to obtain the state ${\cal O}_1(z_1) {\cal O}_2(z_2)|0\rangle$.   This state may be decomposed into irreducible representations of the Virasoro algebra, where as usual we are defining the Hilbert space in terms of radial quantization around the origin $z=0$.    Each such representation is labelled by a corresponding primary field ${\cal O}_p$.  Letting $P_p$ denote the projector onto the representation labelled by ${\cal O}_p$, we write the decomposition as,
\bea
{\cal O}_1(z_1) {\cal O}_2(z_2)|0\rangle = \sum_p C_{12p} {1\over C_{12p}} P_p {\cal O}_1(z_1) {\cal O}_2(z_2)|0\rangle
\eea
where $C_{12p}$ is the primary three-point coefficient, ${\cal O}_1(z){\cal O}_2(0)\sim  C_{12p} z^{h_p-h_1-h_2}{\cal O}_p(0)+ \ldots$.      Each term
\bea
\Psi_{12p}[z_2,z_1] ={1\over C_{12p}}  P_p {\cal O}_1(z_1) {\cal O}_2(z_2)|0\rangle
\eea
 is referred to as an ``OPE block", since it corresponds to using the OPE to express ${\cal O}_1(z_1) {\cal O}_2(z_2)$ in terms of local operators at the origin, and keeping only those operators in the conformal family of ${\cal O}_p$.  Since we have pulled out $C_{12p}$ from its definition, the OPE block is a universal object, completely fixed by Virasoro symmetry.  Its precise form is most efficiently extracted as the solution to Ward identities, as we now discuss.

The CFT stress tensor has the mode expansion,
\bea
T(z) = \sum_{n=-\infty}^\infty l_n z^{-n+2}
\eea
where the $l_n$ obey the Virasoro algebra,
\bea
[l_m,l_n]=(m-n)l_{m+n}+{c\over 12}(m^3-m)\delta_{m,-n}
\eea
For a primary operator ${\cal O}(z)$ of dimension $h$ we have
\bea
[l_n,{\cal O}(z)]= -{\cal L}^{(z)}_n {\cal O}(z)
\eea
where the differential operators,
\bea\label{Ldiff}
\Lc^{(z)}_n = -z^{n+1}\p_{z}-(n+1)h z^n
\eea
obey
\bea
[\Lc^{(z)}_m,\Lc^{(z)}_n]&=(m-n)\Lc^{(z)}_{m+n}
\eea
Using $[l_n,P_p]=0$, along with $l_n|0\rangle =0$ for $n\geq -1$,  and  the trivial identity $[l_n ,{\cal O}_1(z_1) {\cal O}_2(z_2)]  = [l_n ,{\cal O}_1(z_1)] {\cal O}_2(z_2)+{\cal O}_1(z_1)] [l_n,{\cal O}_2(z_2)]$, we have
\bea\label{ward}
\left( l_n +\Lc_n^{(z_2)} + \Lc_n^{(z_1)}\right) \Psi_{12p}[z_2,z_1]=0~,\quad n\geq -1
\eea
We refer to this system of equations as the Virasoro Ward identity.

One approach to determining $\Psi_{12p}[z_2,z_1]$ is to solve (\ref{ward}) order by order in the level expansion.  In particular, take $z_1=0$, $z_2=z$ and expand
\bea
\Psi_{12p}[z,0] =z^{h_p-h_1-h_2} \left( 1+ c_1 zl_{-1}+c_{11}z^2l_{-1}l_{-1}+ c_2 z^2 l_{-2}+ \ldots \right) {\cal O}_p(0) |0\rangle
\eea
The Ward identity turns into recursion relations for the expansion coefficients.   An efficient method for solving these recursion relations is to use an oscillator representation of the $l_n$, as explained in  \cite{Besken:2019bsu}.

We now focus on the ``vacuum OPE block", corresponding to taking ${\cal O}_p $ to be the identity operator $I$.  We further take ${\cal O}_1 = {\cal O}_2 = {\cal O}$, since $C_{12 I}=0$ unless ${\cal O}_1 = {\cal O}_2$.   One way to think about the vacuum OPE block is that it captures all information about correlation functions of ${\cal O }_1(z_1) {\cal O}_2(z_2)$ with any number of stress tensor insertions,
\bea
\langle 0| T(w_1) \ldots T(w_n) {\cal O}_1(z_1) {\cal O}_2(z_2) |0\rangle = \langle 0| T(w_1) \ldots T(w_n) \Psi_{120}[z_2,z_1]
\eea
In the remainder of this section we establish that the unrenormalized Wilson line provides a formal solution to the Ward identity and hence can be identified with the vacuum OPE block,
\bea
\Psi_{120}[z_2,z_1] = W[z_2,z_1]|0\rangle
\eea
As already noted, the solution will be formal in the sense that we will ignore the existence of UV divergences, which is only correct at leading order in $1/c$.
For the same reason, the conformal dimension appearing in the Ward identity will be $h=-j$, corresponding to the large $c$ limit of the general result in (\ref{hj}).

Our objective is to show
\bea\label{wardW}
\left( l_n +\Lc_n^{(z_2)} + \Lc_n^{(z_1)}\right) \langle j,-j|\left[P \exp \int_{z_1}^{z_2} a\right] |j,j\rangle |0\rangle=0~,\quad n\geq -1
\eea
with
\bea\label{adef}
a(z)=L_1+{6\over c}T(z)L_{-1}
\eea
The general strategy is to use the relation between infinitesimal $SL(2,\CC)$ gauge transformations and conformal transformations. Under an infinitesimal $SL(2,\CC)$ gauge transformations we have $\delta_\lambda a= d\lambda +[\lambda,a]$.  We consider
\bea\label{lam}
 \lambda_n(z)  =\eps_n(z)L_1+\p \eps_n(z) L_0+\left({1\over 2} \p^2 \eps_n(z)+{6\over c} T(z)\eps_n(z) \right)L_{-1}
 \eea
 with
 \bea
 \eps_n(z) =z^{n+1}
 \eea
 Note that in (\ref{lam})  $T(z)$ is the stress tensor {\em operator}, so that $\lambda_n(y)$ defines an operator valued gauge transformation.  The form of $\lambda_n(z)$ is chosen so as to preserve the form of $a(z)$ in (\ref{adef}), with $T(z)$ transforming as a stress tensor under the infinitesimal conformal transformation $z\rightarrow z+ \eps_n(z)$,
 \bea
 \delta_{\lambda} T =     \eps_n \p T +2 \p \eps_n T +{c\over 12}\p^3 \eps_n~
 \eea
 Since the Virasoro algebra implies $\delta_{\lambda_n} T(z) = [l_n,T(z)]$, it follows that
 \bea
 \delta_{\lambda_n} a(z) = [l_n,a(z)]~.
 \eea
The gauge covariance of the Wilson line (\ref{gtW}) then yields the relation
\bea\label{gW}
 \left[l_n, P\exp \int_{z_1}^{z_2}a \right]=  \lambda(z_2)\left( P\exp \int_{z_1}^{z_2}a\right) - \left(P\exp \int_{z_1}^{z_2}a  \right)\lambda(z_1)
 \eea

Next, using $L_{-1}|j,j\rangle = \langle j,-j|L_{-1}=0$, we have
\bea
\p_{z_2}  \langle j,-j|\left( P\exp \int_{z_1}^{z_2}a\right)|j,j\rangle &=&  \langle j,-j| L_1 \left( P\exp \int_{z_1}^{z_2}a\right)|j,j\rangle \cr
\p_{z_1}  \langle j,-j|\left( P\exp \int_{z_1}^{z_2}a\right)|j,j\rangle &=& - \langle j,-j|  \left( P\exp \int_{z_1}^{z_2}a\right)L_1|j,j\rangle~.
\eea
We then find
\bea\label{LW}
\Lc_n^{(z_2)}  \langle j,-j|\left( P\exp \int_{z_1}^{z_2}a\right)|j,j\rangle &=&  - \langle j,-j| \lambda_n(z_2)  \left( P\exp \int_{z_1}^{z_2}a\right)|j,j\rangle \cr
\Lc_n^{(z_1)}  \langle j,-j|\left( P\exp \int_{z_1}^{z_2}a\right)|j,j\rangle &=&  \langle j,-j|  \left( P\exp \int_{z_1}^{z_2}a\right)\lambda_n(z_1) |j,j\rangle
\eea
with $\Lc_n^{(z_{1,2})}$ given by the differential operators in (\ref{Ldiff}) with $h=-j$.
Finally, combining (\ref{gW}) and (\ref{LW}) implies the desired Ward identity (\ref{wardW}).

This analysis shows that the Wilson line correctly yields the vacuum Virasoro OPE block in the large-$c$ limit.
In evaluating correlators of the Wilson line with some number of external stress tensors, the leading large-$c$  contribution will not involve any OPE singularities among stress tensors on the Wilson line, and hence no associated divergences, so we have shown that the Wilson line correctly reproduces these leading large-$c$ contributions.   On the other hand, at subleading orders in $1/c$ the Wilson line needs to be renormalized, and the preceding argument does not hold in the presence of a UV regulator.  On general grounds, we expect that once the regulator is taken away, the renormalized Wilson line will obey the Ward identity but with the renormalized dimension (\ref{hj}); evidence for this claim appears in  \cite{Fitzpatrick:2016mtp,Besken:2017fsj,
Hikida:2018dxe,Hikida:2017ehf,Besken:2018zro} as well as in the discussion that follows.


\section{Renormalization of Wilson line correlators}
\setcounter{equation}{0}
\label{sec:4}

In this section, we shall review the regularization and renormalization of the vacuum expectation value of the Wilson line operator in an expansion in powers of $1/c$. We shall summarize the results of \cite{Besken:2018zro}
which prove that this correlator indeed obeys the form predicted for a bi-local operator of dimension $h(j)$ given in (\ref{Wconf}) to order $1/c^3$. We shall then present new results for the correlator of the Wilson line with one stress tensor insertion and show that it obeys (\ref{TWconf}) to order $1/c$. Finally, we present arguments that the equivalence of (\ref{al}) holds to leading order in $1/c$ for an arbitrary number of stress tensor insertions.

\subsection{Regularization}

No regularization of short distance singularities which preserves the full conformal symmetry in two-dimensional space-time is known to exist. However, dimensional regularization  from two dimensions to $d=2-\ep$ dimensions preserves dilation symmetry of stress tensor correlators in $d$ dimensions for all values of $d$ where the integrals of the individual Feynman graphs are absolutely convergent. For this reason, dimensional regularization and subsequent analytic continuation in $\ep$ appears better-suited for regularizing correlators in scale invariant theories than other schemes.\footnote{The use of short-distance regulators other than dimensional regularization was attempted in \cite{Besken:2017fsj} with a dimensionful cut-off and in section~6 of \cite{Besken:2018zro} with a strictly two-dimensional regularization of the operator. Both lead to inconsistency with the conformal Ward identities at sufficiently high order in $1/c$.}

\sm

As should be expected in $d \not=2$ the Ward identity (\ref{TT}), by which all stress tensor correlators can be computed on the two-dimensional plane, no longer holds and cannot be used to this end. To evaluate stress tensor correlators in the absence of the Ward identities a concrete quantum field theory model is needed which is valid in arbitrary dimension $d$. Such a model is provided by taking $c$ integer and considering $c$ scalar fields $\phi ^\gamma$ with $\gamma =1, \cdots, c$ in $d$ space-time dimensions. After renormalization and continuation to two dimensions, it should be expected that all correlators become independent of the specific model used. The fact that $c$ takes integer values is immaterial in the large $c$ expansion used here.

\sm

The free-field correlators in this model are standard.  Parametrizing space-time $\RR^d$ by coordinates $(z, \bar z,  \bz)$ where $z, \bar z$ are the complex coordinates for $\CC$ and $\bz \in \RR^{d-2}$, we readily evaluate the normalized two-point function of the field $\p_z \phi^\gamma$,
\bea
\label{prop1}
\left \< \p_z \phi ^{\gamma}(z) \p_w \phi ^{\gamma '} (w)  \right \> = { -  \delta ^{\gamma \gamma '} (\bar z- \bar w)^2
\over \big ( |z-w|^2 +(\bz - \bw)^2 \big )^{{d \over 2} +1}}
\eea
The overall normalization may be absorbed by a renormalization function in the Wilson line and is immaterial, as we shall see below. For two points in the complex plane we have $\bz=\bw=0$, and for two points on the real line the correlator in $d=2-\ep$ dimensions  simplifies to the following formula we shall use throughout,
\bea
\label{prop2}
\left \< \p_w \phi ^{\gamma}(z) \p_z \phi ^{\gamma '} (w)  \right \>
= { -   \delta ^{\gamma \gamma '}   \over  |z-w|^{2-\ep}}
\eea
In this model, the holomorphic stress tensor $T(z)$ for $z \in \CC$ is defined as the $T_{zz}$ component of the $d$-dimensional traceless stress tensor for the free field $\phi^\gamma$, which is given by,
\bea
T(z) = - \half \sum _{\gamma=1}^c : \p_z \phi ^\gamma (z) \p_z \phi^\gamma (z) :
\eea
where the normal ordering symbol $::$ instructs us to omit all self-contractions.

\subsection{Renormalization}

Following the discussion of the preceding subsection, we shall adopt the model of $c$ free scalar fields in dimension $d=2-\ep$ as a systematic regularization, and evaluate all stress tensor correlators with the help of (\ref{prop1}). We shall translate to $z_1=0$, rotate to $z_2=z \in \RR^+$, and choose the path of integration  along $\RR^+$. Additional stress tensors will be inserted at points $w_1, \cdots, w_p \in \RR^-$ in (\ref{al}), and subsequently analytically continued to the complex plane. Therefore, all correlators may effectively be evaluated by using (\ref{prop2}).

\sm

We know of no systematic procedure to renormalize a highly composite operator such as the Wilson line and to guarantee that its renormalized correlators satisfy the two-dimensional conformal Ward identities. The proposal made in \cite{Besken:2018zro} for the renormalized Wilson line is to include two renormalization functions: a function $N$ multiplying the  Wilson line operator, and a function $\alpha$ multiplying  $1/c$,
\bea
\label{W}
W _R [z,0] = N  \< j, -j |
P \exp \left \{ \int _0 ^z dy  \Big (L_1 + { 6  \alpha   \over c}  T(y) L_{-1} \Big ) \right \} |j,j\>
\eea
The functions $N$ and $\alpha$ depend on the regulator $\ep$ as well as on $c$ and $j$.
Below we shall summarize the results of \cite{Besken:2018zro} which show that, in an expansion in powers of $1/c$, these two renormalization functions suffice to the vacuum expectation value $\< W_R[z,0]\>$ to order $1/c^3$ included. Whether $N$ and $\alpha$ suffice to renormalize $W_R$ to all orders in $1/c$, or non-perturbatively in $c$ remains an open question.

\subsection{Expansion in powers of $1/c$}
\label{sec:43}

To carry out a systematic expansion in powers of $1/c$, it is useful to recast the Wilson line operator in the ``interaction picture", by using the following identity,
\bea
P \exp \left \{ \int _0 ^z dy  \Big (L_1 + { 6  \alpha   \over c}  T(y) L_{-1} \Big ) \right \}
= e^{zL_1} P \exp \left \{ { 6  \alpha  \over c} \int _0 ^z dy \, X(y) \, T(y) \right \}
\eea
where $X(y)$ is given by,
\bea
X(y) = e^{-y L_1} L_{-1} e^{y L_1} = L_{-1} - 2 yL_0 + y^2 L_1
\eea
The Wilson line operator thus takes the form,
\bea
\label{WX}
W _R [z,0] = N  \< j, -j |
e^{zL_1} P \exp \left \{ { 6  \alpha  \over c} \int _0 ^z dy \, X(y) \, T(y) \right \}  |j,j\>
\eea
The expansion in powers of $1/c$ is obtained by first expanding in powers of $\alpha/c$, then constructing $N$ and $\alpha$ in a power series in $1/c$, and finally extracting from this combination the systematic expansion in $1/c$. The expansion in $\alpha/c$ of a general correlator, is given by,
\bea
\label{TpWn1}
\< T(w_1) \cdots T(w_p) W_R[0,z] \>
=
N z^{2j-2p} \sum _{n=0} ^\infty  {  \alpha^n \over c^n} \, z^{(n+p)\ep}  T^p W_n
\eea
where the coefficients $T^pW_n$ are given by,
\bea
T^p W_n = {6^n \, z^{2p} \over z^{(n+p)\ep}} \int _0 ^ z dy_n \cdots \int _0 ^{y_2} dy_1
F_n (z;y_n, \cdots , y_1) \<  T(w_1) \cdots T(w_p) T(y_1) \cdots T(y_n)\>
\eea
and the functions $F_n$ are defined by,
\bea
F_n(z;y_n, \cdots, y_1) = \< j, -j| X(y_n) \cdots X(y_1) |j,j\>
\eea
The function $F_n(z;y_n, \cdots, y_1)$ is a homogeneous polynomial  of degree $n$ in its variables $z, y_n, \cdots , y_1$. The factor $z^{2p-(n+p)\ep}$ is included in the coefficients $T^pW_n$ to render it scale-invariant and thus independent of $z$ when expressed in terms of the dimensionless variables $v_1=w_1/z, \cdots, v_p=w_p/z$ and $x_1=y_1/z, \cdots , x_n=y_n/z$, so that we have,
\bea
\label{TpWn2}
T^p W_n = 6^n  \int _0 ^ 1 dx_n \cdots \int _0 ^{x_2} dx_1
F_n (1;x_n, \cdots , x_1) \<  T(v_1) \cdots T(v_p) T(x_1) \cdots T(x_n)\>
\eea
The functions $F_n$ obey a recursion relation, which allows for their ready evaluation \cite{Besken:2018zro}.

\sm

The renormalization functions $N$ and $\alpha$ may similarly be expanded in a power series in $1/c$,
\bea
N= 1 + \sum _{n=1}^\infty { N_n \over c^n}
\hskip 1in
\alpha = 1 + \sum _{n=1}^\infty { \alpha _n \over c^n}
\eea
where the coefficients $N_n$ and $\alpha _n$ depend on $j$ and have a Laurent expansion in powers of $\ep$.
They are expected to be independent of the external stress tensors, and thus independent of the number $p$.
The most singular power of $\ep$ in $N_n$ and $\alpha_n$ is expected to be $1/\ep^n$.

\subsection{Regularized stress tensor correlators}

The stress tensor correlators are evaluated in the theory of $c$ free scalar fields. We shall place the points $w_1,\cdots, w_p$ on $\RR^-$ so that the stress tensor correlators may be computed with two-point function (\ref{prop2}). The combinatorics of these correlators was given in detail in \cite{Besken:2018zro}, and proceeds as follows.
Evidently, we have $\< T(y)\>=0$. The Feynman diagrams for a correlator $\< T(y_1) \cdots T(y_m)\>$ for $m\geq 2$ may be distinguished by  the number of connected  one-loop sub-graphs.  Each sub-graph  may be labelled by a {\sl partition $P$ into cycles} of the set of points $\{ y_1, \cdots, y_m\}$, with  each cycle  containing at least two points. Two partitions are equivalent if they are related by cyclic permutations and/or reversal of orientation of the points in each cycle, and under permutations of the cycles.  This partitioning of a Feynman graph  into inequivalent cycles is unique.

\sm

We shall denote a cycle of ordered points $y_{i_1} , \cdots , y_{i_\mu}$ by a square bracket $[i_1, \cdots, i_\mu]$
and the value of the corresponding one-loop diagram along this cycle by,
\bea
\label{T123}
\< T^2 \> _{[i_1, i_2]} & = & { c/2 \over |y_{i_1} - y_{i_2}|^{4-2\ep}}
\no \\
\< T^\mu \> _{[i_1, \cdots, i_\mu]} & = & { c \over |y_{i_1} - y_{i_2}|^{2-\ep} |y_{i_2} - y_{i_3}|^{2-\ep}
\cdots |y_{i_\mu} - y_{i_1}|^{2-\ep} }\, , \hskip 0.5in \mu \geq 3
\eea
The correlator  is given by a sum over all possible inequivalent partitions $P= C_1 \cup C_2 \cup \cdots \cup C_\nu$ into $\nu$ cycles, with $C_s \cap C_{s'} = \emptyset$ for $s' \not= s$, of the set $\{ y_1, \cdots, y_m\}$,
\bea
\label{TC}
\big \< T(y_1) \cdots T(y_m) \big \> = \sum _P \< T^m\> _P~,
\hskip 1in
\< T^m\> _P= \prod _{s=1}^\nu \< T^{\mu_s} \> _{C_s}
\eea
For example, when $m=4$ six partitions contribute,
\bea
\label{tdec}
\<T(y_1) \cdots T(y_4)\> & = & \< T^4\> _{[12][34]}+ \< T^4\> _{[13][24]}+ \< T^4\> _{[14][23]}
\no \\ &&
+ \< T^4\> _{[1234]}+ \< T^4\> _{[1342]}+ \< T^4\> _{[1324]}
\eea
When additional stress tensors $T(w_1) \cdots T(w_p)$ are present, the above combinatorics should be applied to the set of all $m=n+p$ stress tensors.

\subsection{Renormalization of $\< W_R[z,0]\>$}

It was shown in \cite{Besken:2018zro} that the functions $N$ and $\alpha$ suffice to renormalize $\< W_R [z,0]\>$, produce the scaling in $z$ of (\ref{Wconf}), with chiral dimension of (\ref{hj}) to order $1/c^3$,
\bea
\label{hjj}
h(j) = -j -{6 \over c} j(j+1) - {78 \over c^2} j(j+1) - { 1230 \over c^3} j(j+1) + \cO(c^{-4})
\eea
provided the renormalization function $\alpha$ is chosen as follows,
\bea
\label{alpha}
\alpha    = 1 + {1 \over c} \left ( {6 \over \ep} + 3   \right )
+{1 \over c^2} \left ( { 30 \over \ep^2} + {55 \over \ep} +  {185 \over 3}  - { 1 6 \pi^2 \over 5} \big ( 3 j(j+1) -1 \big )  \right ) +\cO(c^{-3})
\eea
Terms of order $1/c^3$ are not needed to evaluate $h(j)$ to the order given in (\ref{hjj}). The expansion of the function $N$ to order $1/c^3$ included is required to evaluate (\ref{hjj}) but is significantly more involved than $\alpha$, and we shall give its expression to order $1/c^2$ only,
\bea
\label{N}
N & = &  1 -   {6J^2  \over c \, \ep} -  { 1 \over c}  (10J^2 -6j  )
+  {6J^2 \over c^2 \, \ep^2} (3J^2 -2)
 + { 3 J^2 \over c^2 \, \ep} (20J^2-16j-17)
 \no \\ && \hskip 0.6in
+ { 1 \over c^2} \left ( (50j^2+2j - 160)J^2  +42j +8 \pi^2J^2   \right ) + \cO(c^{-3})
\eea
where $J^2=j(j+1)$ is the quadratic Casimir value in the representation with spin $j$.


\section{Renormalization and bi-locality of $\< T(w) W_R[z,0]\>$}
\setcounter{equation}{0}
\label{sec:5}

If a renormalized Wilson line operator can be constructed which satisfies the criteria of bi-locality and conformal behavior spelled out in subsection \ref{sec:25}, then its correlator with one stress tensor insertion must behave according to (\ref{TWconf}),
\bea
\label{TWconf2}
 \< T(w) W_R[z,0] \> = { h(j) \, z^2 \over w^2 (w-z)^2} \, \< W_R[z,0]\>
\eea
In the absence of complete arguments for the existence of such an operator we construct its correlators  in a power expansion in $1/c$, and  verify that they satisfy the criteria spelled out in subsection \ref{sec:25}. In this section, we shall do so for the correlator $\< T(w) W_R[z,0]\>$ to order  $1/c$, which is the lowest order at which non-trivial renormalization is required.

\subsection{Expansion in $1/c$}

To compute $ \< T(w) W_R[z,0] \>$ we use the formalism of subsection \ref{sec:43} for $p=1$ which gives,
\bea
 \< T(w) W_R[z,0] \> = N z^{2j-2} \sum _{n=0}^\infty { \alpha ^n \over c^n} z^{(n+1)\ep} TW_n
 \eea
 where the coefficients $TW_n$ are given in terms of the scaling variable $v=w/z \in \RR^-$ by,
 \bea
 TW_n = 6^n \int _0^1 dx_n \cdots \int _0 ^{x_2} dx_1 F_n (1;x_n, \cdots, x_1) \< T(v) T(x_1) \cdots T(x_n)\>
 \eea
The renormalization functions $N$ and $\alpha$ are identical to those derived in (\ref{N}) and (\ref{alpha}) for the renormalization of $\< W_R [z,0]\>$ which, to this order, are given by their contributions to order $1/c$ included.
We shall study the correlator to order $1/c$ and thus retain contributions from $n=1,2,3$ (since $\< T(v)\>=0$), which requires the following stress tensor correlators,
\bea
\label{T123}
\< T(v) T(x_1)\> & = & { c/2 \over (x_1 - v)^{4-2\ep}}
\no \\
\< T(v) T(x_1) T(x_2) \> & = & { c \over (x_1-v)^{2-\ep}  (x_2-v)^{2-\ep}  (x_2-x_1)^{2-\ep} }
\no \\
\< T(v) T(x_1) T(x_2) T(x_3) \> & = &
{ c^2/4 \over (x_1-v)^{4-2 \ep}  (x_3-x_2)^{4-2\ep} } + 2 \hbox{ perms}
 + \cO(c)
\eea
The $\cO(c)$ terms in the four-$T$ correlator will not contribute to order $1/c$. The contributions from $n=1,2$ are given by the integrals,
\bea
\label{n=12}
TW_1 & = & 3  \int _0 ^1 dx_1 \,  { F_1(1;x_1) \over (x_1-v)^{4-2\ep}}
\no \\
TW_2 & = & {36 \over c}  \int _0 ^1 dx_2 \int _0 ^{x_2} dx_1 \,
{ F_2(1;x_2,x_1) \over (x_1-v)^{2-\ep} (x_2-v)^{2-\ep} (x_2-x_1)^{2-\ep} }
\eea
The contribution from $n=3$ is the sum of three terms given by the three terms in the corresponding four-$T$ correlator of (\ref{T123}),
\bea
TW_3 = TW_3^{(1)} + TW_3^{(2)} + TW_3^{(3)}
\eea
with each term given by,
\bea
TW_1^{(1)}  & = & {54 \over c}   \int _0 ^1 dx_3 \int _0 ^{x_3} dx_2  \int _0 ^{x_2} dx_1 \,
{ F_3(1;x_3,x_2,x_1)  \over (x_1-v)^{4-2\ep} (x_3-x_2)^{4-2\ep} }
\no \\
TW_2^{(2)}  & = & {54 \over  c}   \int _0 ^1 dx_3 \int _0 ^{x_3} dx_2  \int _0 ^{x_2} dx_1 \,
{ F_3(1;x_3,x_2,x_1)  \over (x_2-v)^{4-2\ep} (x_3-x_1)^{4-2\ep} }
\no \\
TW_3^{(3)}  & = & {54 \over c}   \int _0 ^1 dx_3 \int _0 ^{x_3} dx_2  \int _0 ^{x_2} dx_1 \,
{ F_3(1;x_3,x_2,x_1)  \over (x_3-v)^{4-2\ep} (x_2-x_1)^{4-2\ep} }
\eea
The functions $F_n$ are given by,
\bea
F_1 (1;x_1)& = & 2jx_1(x_1-1)
\\
F_2 (1;x_2,x_1) & = & 2jx_1(x_2-1)  \big ( (2j-1) x_1x_2  -2j x_2+x_1 \big )
\no \\
F_3 (1;x_3,x_2,x_1) & = & 4j x_1 (x_3-1) \Big ( (2j^2-3j+1)x_1x_2^2x_3 - (2j^2-j) x_2^2x_3 + (2j-1) x_1 x_2^2
\no \\ && \quad
 - j x_2^2
-(2j^2-3j+1) x_1x_2x_3 + 2 j^2 x_2x_3 -(j-1) x_1x_2 -j x_1 x_3 \Big )
\no
\eea
To evaluate the contributions of order $c^0$ and $c^{-1}$ to $\< T(w) W_R[z,0]\>$, we need $TW_1$ to orders $\ep^0$ and $\ep^1$ since the renormalization $ N\alpha $, which multiplies this contribution, will pick up order $\ep^0$ terms that are of order $1/c$. The remaining terms are needed to orders $\ep^{-1}$ and $\ep^0$.

\subsection{Calculation of the coefficients $TW_n$}

The integral for $TW_1$ is straightforward and we find,
\bea
TW_1 & = &  { - j \over v^2 (v-1)^2}  - 2 j \ep \left (  {1 + 2v \over v^2} \ln (-v) + { 3 -2v \over (v-1)^2}  \ln (1-v) \right .
\no \\ && \hskip 1.5in \left .
+ {2 \over v(1-v)} + {5 \over 6v^2(v-1)^2} \right  )  + \cO(\ep^2)
\eea
The expression is symmetric under $v \to 1-v$.

\sm

To evaluate $TW_2$ we set $x_2=x$ and change variables from $x_1$ to $x_1=x-u$ with $0 \leq u \leq x$,
\bea
TW_2 = {36 \over c}  \int _0 ^1 dx \int _0 ^x du \,
{ F_2(1;x,x-u) \over (x-u-v)^{2-\ep} (x-v)^{2-\ep} u^{2-\ep} }
\eea
Next, we define the function,
\bea
R_\ep = { 1 \over (x-u-v)^{2-\ep}} - { 1 \over (x-v)^{2-\ep}} - { (2-\ep) u \over (x-v)^{3-\ep}}
\eea
which is constructed so as to vanish to second order in $u$ at $u=0$ for all values of $\ep$. In terms of $R_\ep$, we may recast $TW_2$ as follows,
\bea
TW_2 = {36 \over c}  \int _0 ^1 dx \int _0 ^x du \,
{ F_2(1;x,x-u) \over (x-v)^{2-\ep} u^{2-\ep}}
\left ( { 1 \over (x-v)^{2-\ep}} + { (2-\ep) u \over (x-v)^{3-\ep}} + R_\ep \right )
 \eea
Since $R_\ep$ vanishes as $u^2$, the integral over $R_\ep$ converges for $\ep=0$, so that we have,
\bea
TW_2 & = & {36 \over c}  \int _0 ^1 dx \int _0 ^x du \,
{ F_2(1;x,x-u) \over (x-v)^{5-2\ep} \, u^{2-\ep}} \,  ( x-v + (2-\ep) u )
\no \\ && +
 {36 \over c}   \int _0 ^1 dx \int _0 ^x du \,
 {F_2(1;x,x-u) \over (x-v)^2 u^2}   R_0  + \cO(\ep)
 \eea
Since $F_2$ is polynomial in its arguments, each integral is elementary and may now be performed using MAPLE: the result is quite involved and will not be exhibited here.

\sm

To evaluate $TW_3^{(i)}$ for $i=1,2,3$, we rearrange the integrations so as to carry out the intgerals over the corresponding $x_i$ last. To this end, we  relabel $(x_1,x_2,x_3) \to (x_3,x_2,x_1)$ in $TW_3^{(1)} $, and $(x_1,x_2,x_3) \to (x_1,x_3,x_2)$ in $TW_2^{(2)} $ leaving the last integral unchanged. Next, we also change variables from $x_1 $ to $u=x_2-x_1$, so that we find,
\bea
\label{TWF123a}
TW_1^{(1)}  & = & {54 \over c}   \int _0 ^1 dx_3 \int _{x_3} ^1 dx_2  \int _{0} ^{x_2-x_3}  du \,
{ F_3(1;x_2,x_2-u,x_3)  \over (x_3-v)^{4-2\ep} \, u^{4-2\ep} }
\no \\
TW_2^{(2)}  & = & {54 \over  c}  \int _0 ^1 dx_3 \int _{x_3} ^1 dx_2  \int _{x_3-x_2} ^{x_2} du \,
{ F_3(1;x_2,x_3,x_2-u)  \over (x_3-v)^{4-2\ep} \, u^{4-2\ep} }
\no \\
TW_3^{(3)}  & = & {54 \over c}  \int _0 ^1 dx_3 \int _0 ^{x_3} dx_2  \int _0 ^{x_2} du \,
{ F_3(1;x_3,x_2,x_2-u)  \over (x_3-v)^{4-2\ep} \, u^{4-2\ep} }
\eea
Interchanging the integrations over $x_2$ and $u$ reveals the convenient fact that the integrals over $x_2$ are of a polynomial in $x_2$ and may be performed easily, whereafter the integrals over $u$ and $x_3$ may be performed in MAPLE by elementary methods as well. Again, the results are quite involved and will not be presented individually here.

\sm

Assembling all contributions, inserting the renormalization functions $N$ and $\alpha $, and using MAPLE to simplify the final results, we find that all logarithmic contributions $\ln (-v) = \ln (-w/z)$ and $\ln (1-v)=\ln (1-w/z)$ cancel exactly to this order, and the resulting correlator is governed by the relation (\ref{TWconf2}) expected under the assumption of conformal invariance with the operator dimension  $h(j)$ given by (\ref{hjj}) to order $\cO(1/c)$ included.


\section{Higher order  correlators $\< T(w_1) \cdots T(w_p) W_R [z,0]\>$}
\setcounter{equation}{0}
\label{sec:6}

Assuming the existence of a bi-local conformal primary renormalized Wilson line operator $W_R[z,0]$, its correlator  with $p$ stress tensor insertions can be derived completely from the conformal Ward identities.  In turn, showing that such correlators obey the relations expected from conformal invariance and bi-locality will provide further evidence for the existence of a renormalized Wilson line operator satisfying the criteria of subsection \ref{sec:25}.

\sm

In this last section, we shall show that even to leading order in $1/c$, where no renormalization effects are required, individual contributions to the correlator do not respect conformal properties, and in particular exhibit logarithmic singularities when $z$ approaches one of the insertion points $w_1, \cdots, w_p$. For the correlator with $p=1$, this issue did not arise as the only contribution was given by the coefficient $TW_1$ whose conformal behavior is manifest. We shall begin by examining the simplest case, namely for $p=2$, and show by explicit calculation that the combined contributions to the correlator $\< T(w_1) T(w_2) W_R [z,0]\>$ see all their logarithmic contributions at $z\sim w_1, w_2$  cancelled to order $c^0$. We shall then present general arguments why this result extends to all values of $p$ at order $c^0$.

\subsection{The correlator $\< T(w_1) T(w_2) W_R [z,0]\>$}

The correlator is given by (\ref{TpWn1}) and (\ref{TpWn2}) for the case $p=2$. To order $c^0$, no short-distance singularities appear, we may set $N=\alpha =1$, and $\ep=0$. The contribution from $n=0$ is given by a disconnected correlator which is of order $c$ and trivial. The remaining connected contributions of order $c^0$ are given by the $n=1,2$. Working with rescaled variables $v=w/z, x=y/z$, we need the three-$T$ stress tensor correlators,
\bea
\label{T123}
\< T(v_1) T(v_2) T(x_1)\>  =  { c \over (x_1 - v_1)^2 (x_1 - v_2)^2 (v_2 - v_1)^2}
\eea
and the four-$T$ connected correlators,
\bea
\< T(v_1) T(v_2) T(x_1) T(x_2) \>_c  =
{ c^2/4 \over (x_1-v_1)^4 (x_2-v_2)^4 }
+  { c^2/4 \over (x_1-v_2)^4 (x_2-v_1)^4 }
\eea
The corresponding contributions $T^2W_1$ and $T^2W_2$, to order $c^0$, are given by,
\bea
T^2W_1 & = &  \int _0^1 dx_1 {  6 F_1(1;x_1) \over (x_1-v_1)^2 (x_1-v_2)^2 (v_2-v_1)^2}
\no \\
T^2W_2 & = &
  \int _0^1 dx_2 \int ^{x_2} _0 d x_1    \left ( {9 F_2(1;x_2,x_1)  \over (x_1-v_1)^4 (x_2-v_2)^4 } +
{9 F_2(1;x_2,x_1)    \over (x_1-v_2)^4 (x_2-v_1)^4 } \right )
\eea
Evaluation of the integrals is elementary, and we find for $T^2W_1$,
\bea
T^2W_1  =  { 24 j \over (v_1-v_2)^4 } + { 12 j (2v_1v_2-v_1-v_2) \over (v_1-v_2)^5}
\ln \left ( { v_1(1-v_2) \over v_2(1-v_1)} \right )
\eea
The contribution $T^2W_2$ similarly has logarithmic singularities which, however, all cancel in the sum with $T^2W_1$, and we find,
\bea
T^2W_1 +  T^2W_2  =  { j^2 \over v_1^2 (1-v_1)^2  v_2^2 (1-v_2)^2 }
+ { j \over v_1 (1-v_1) v_2 (1-v_2) (v_1-v_2)^2}
\eea
Assembling all contributions and reverting to the variables $w$ and $z$, we find,
\bea
\< T(w_1)  T(w_2) W_R [z,0]\>
={ j^2 z^{2j+4} \over w_1^2 (z-w_1)^2  w_2^2 (z-w_2)^2 }
+ { j z^{2j+2} \over w_1 (z-w_1) w_2 (z-w_2) (w_1-w_2)^2}
\eea
in agreement with the predictions from conformal Ward identities.

\subsection{The correlator $\langle T(w_1)\ldots T(w_n) W_R[z,0]\rangle$ at leading order in $1/c$}

In section \ref{wardsec} we derived the Ward identity
\bea
\label{Wident}
\left(l_n +{\cal L}_n^{(z_2)} + {\cal L}_n^{(z_1)}\right) W[z_2,z_1]|0 \rangle =0~,\quad n\geq -1
\eea
at leading order in the $1/c$ expansion where renormalization is not necessary.   We conclude by showing how the Ward identity fixes correlators of the Wilson line with any number of external stress tensor insertions.   We proceed recursively.  Given $\langle T(w_1)\ldots T(w_n) W[z_2,z_1]\rangle$ we wish to insert an additional stress tensor.   We work in the operator formulation.  Using the general structure of the Virasoro algebra we have $[l_n,T]\sim T + I $, so we can write
\bea
\langle T(u)  T(w_1)\ldots T(w_n) W[z_2,z_1]\rangle &=&
  \sum_{m=2}^\infty {1\over u^{m+2}} \langle l_m  T(w_1)\ldots T(w_n) W[z_2,z_1]\rangle \cr
&=& \sum_{m=2}^\infty {1\over u^{m+2}} \langle T(w_1)\ldots T(w_n) l_m  W[z_2,z_1]\rangle +{\rm (fewer)}\cr
&=& - \sum_{m=2}^\infty {1\over u^{m+2}} \left({\cal L}_m^{(z_2)} + {\cal L}_m^{(z_1)} \right)\langle T(w_1)\ldots T(w_n)  W[z_2,z_1]\rangle +{\rm (fewer)}\cr
&&
\eea
where ``fewer" stands for correlators with $n$ or $n-1$ stress tensor insertions, which are assumed to be known.  These recursion relations determine all correlators of the Wilson line with stress tensors in terms of $\langle W[z_2,z_1]\rangle$.

Our expectation is that the quantum Wilson line, renormalized in the way we have discussed, will continue to obey (\ref{Wident}) but with the renormalized value of $h$ appearing in ${\cal L}_n$.  If so, this will  fix all correlators with any number of stress tensor insertions in terms of  $\langle W_R[z_2,z_1]\rangle$.

\subsection*{Acknowledgments}

We thank Mert Besken and  Ashwin Hegde for collaboration on the work reviewed here.


\end{document}